\documentclass[prb,showpacs,aps,twocolumn,superscriptaddress]{revtex4-1}
\usepackage{graphicx}% Include figure files
\usepackage{epstopdf}
\usepackage{bm}% bold math
\usepackage{color}
\usepackage{ulem}
\usepackage{amsfonts,amsthm}
\usepackage{amsmath}
\usepackage{amssymb}

\begin{document}
\newcommand{\tr}[1]{\textcolor{red}{#1}}
\newcommand{\tm}[1]{\textcolor{magenta}{#1}}
\newcommand{\tms}[1]{\textcolor{magenta}{\sout{#1}}}
\newcommand{\tb}[1]{\textcolor{blue}{#1}}
\newcommand{\avrg}[1]{\left\langle #1 \right\rangle}
\newcommand{\eqsa}[1]{\begin{eqnarray} #1 \end{eqnarray}}
\newcommand{\eqwd}[1]{\begin{widetext}\begin{eqnarray} #1 \end{eqnarray}\end{widetext}}
\newcommand{\hatd}[2]{\hat{ #1 }^{\dagger}_{ #2 }}
\newcommand{\hatn}[2]{\hat{ #1 }^{\ }_{ #2 }}
\newcommand{\wdtd}[2]{\widetilde{ #1 }^{\dagger}_{ #2 }}
\newcommand{\wdtn}[2]{\widetilde{ #1 }^{\ }_{ #2 }}
\newcommand{\cond}[1]{\overline{ #1 }_{0}}
\newcommand{\conp}[2]{\overline{ #1 }_{0#2}}
\newcommand{\nn}{\nonumber\\}
\newcommand{\cdt}{$\cdot$}
\newcommand{\bra}[1]{\langle#1|}
\newcommand{\ket}[1]{|#1\rangle}
\newcommand{\braket}[2]{\langle #1 | #2 \rangle}
\newcommand{\bvec}[1]{\mbox{\boldmath$#1$}}
\newcommand{\blue}[1]{{#1}}
\newcommand{\bl}[1]{{#1}}
\newcommand{\bn}[1]{\textcolor{blue}{#1}}
\newcommand{\rr}[1]{{#1}}
\newcommand{\bu}[1]{\textcolor{blue}{#1}}
\newcommand{\red}[1]{\textcolor{red}{#1}}
\newcommand{\fj}[1]{{#1}}
\newcommand{\green}[1]{{#1}}
\newcommand{\gr}[1]{\textcolor{green}{#1}}
\newcommand{\cyan}[1]{\textcolor{cyan}{#1}}
\newcommand{\grs}[1]{\textcolor{green}{\sout{#1}}}
\newcommand{\fry}[1]{\fcolorbox{red}{yellow}{#1}}
\newcommand{\frys}[1]{\fcolorbox{red}{yellow}{\sout{#1}}}
\newcommand{\fy}[1]{\fcolorbox{white}{yellow}{#1}}
\newcommand{\fys}[1]{\fcolorbox{white}{yellow}{\sout{#1}}}
\newcommand{\fb}[1]{\tm{\framebox[8.5cm]{\parbox{8.4cm}{#1}}}} 
\newcommand{\fbs}[1]{\tm{\framebox[8.5cm]{\parbox{8.4cm}{\sout{#1}}}}} 
\definecolor{green}{rgb}{0,0.5,0.1}
\definecolor{blue}{rgb}{0,0,0.8}
\preprint{APS/123-QED}

\title{
Stabilization of Topological Insulator Emerging from Electron Correlations
\\ on Honeycomb Lattice and Its Possible Relevance in Twisted Bilayer Graphene}
%\if0
\author{Moyuru Kurita}
\affiliation{Department of Applied Physics, The University of Tokyo, Hongo, Bunkyo-ku, Tokyo, 113-8656, Japan}
\author{Youhei Yamaji}
\affiliation{Quantum-Phase Electronics Center (QPEC), The University of Tokyo, Hongo, Bunkyo-ku, Tokyo, 113-8656, Japan}
\author{Masatoshi Imada}
\affiliation{Department of Applied Physics, The University of Tokyo, Hongo, Bunkyo-ku, Tokyo, 113-8656, Japan}
%\affiliation{CREST, JST, Hongo, Bunkyo-ku, Tokyo, 113-8656, Japan.}
%\date{February 25, 2014}% It is always \today, today,
\date{\today}% It is always \today, today,

\begin{abstract}
Realization and design of topological insulators emerging from electron correlations, called topological Mott insulators (TMIs),
is pursued by using mean-field approximations as well as multi-variable variational Monte Carlo (MVMC) methods  for Dirac electrons on honeycomb lattices.
The topological insulator phases predicted in the previous studies by the mean-field approximation for an extended Hubbard model on the honeycomb lattice turn out to disappear, when we consider the possibility of a long-period charge-density-wave (CDW) order taking over the TMI phase.
Nevertheless, we further show that the TMI phase is still stabilized when we are able to tune the Fermi velocity of the Dirac point of the electron band.
Beyond the limitation of the mean-field calculation, we apply the newly developed MVMC to make accurate predictions after including the many-body and quantum fluctuations.
By taking the extrapolation to the thermodynamic and weak external field limit, we present realistic criteria for the emergence of the topological insulator caused by the electron correlations.
By suppressing the Fermi velocity to a tenth of that of the original honeycomb lattice, the topological insulator emerges in an extended region as a spontaneous symmetry breaking surviving competitions with other orders.
We discuss experimental ways to realize it in a bilayer graphenesystem.
\end{abstract}

\pacs{05.30.Rt,71.10.Fd,73.43.Lp,71.27.+a}% PACS, the Physics and Astronomy
%05.30.Rt Quantum phase transition
%71.10.Fd Hubbard model electronic structure
%73.43.Lp quantum Hall effects
%71.27.+a Strongly correlated electron systems

\maketitle

\section{Introduction}
Recently, topological properties of time-reversal-invariant band insulators in two and three dimensions have been extensively studied\cite{Kane1, Kane2, Fu, Moore, Roy, Hasan}.
%\gr{[Topological invariants should be introduced here !]}
A class of
insulators preserving the time reversal symmetry
is called topological insulators characterized by non-trivial topological invariants\cite{Kane1, Kane2, Fu}.%\gr{[cite Kane-Mele and Fu-Kane-Mele]}.
The topological insulators have been intensively studied because of the existence and potential applications of robust surface metallic states.

Both in two and three dimensions, the topological phases are typically realized in the systems with strong spin-orbit interaction\cite{Bernevig, Konig, Hsieh}.
All the known topological insulators contain heavy or rare metal elements, such as bismuth or iridium,
which poses constraints on the search for topological materials.

Irrespective of constitutents, ubiquitous mutual Coulomb repulsions among electrons have been proposed to generate
effective spin-orbit couplings~\cite{Raghu,Wen, Zhang, Kurita2}.
It has been proposed that an extended Hubbard model on the honeycomb lattice can generate an effective spin-orbit interaction from a spontaneous symmetry breaking at the Hartree-Fock mean-field level leading to a topologically non-trivial phase\cite{Raghu,Kurita2}.
Since the honeycomb-lattice system, which is Dirac semimetals in the non-interacting limit, becomes a topologically nontrivial insulator driven by the Coulomb interaction, this phase is often called a topological Mott insulator (TMI).
This phenomenon is quite unusual not only because an emergent spin-orbit interaction appears from the electronic mutual Coulomb interaction, but also it shows an unconventional
quantum criticality that depends on the electron band dispersion near the Fermi point\cite{Kurita}.

However, this proposed topological phase by utilizing the ubiquitous Coulomb repulsions %\gr{\sout{the idea of TMI}[the words sounds like``the TMI point of view"]}
has not been achieved in real materials even though the TMI is proposed not only in
various solids~\cite{Raghu,Wen, Zhang,Kurita2,Herbut3}
but also in cold atoms loaded in optical lattices~\cite{Kitamura}.
%\cite{Wen, Zhang, Kurita2}.
Even in simple theoretical models such as extended Hubbard models, it is not clear whether the TMIs become stable against competitions with other orders and quantum fluctuations.

Reliable examination of stable topological Mott orders in the extended Hubbard model is
hampered by competing symmetry breakings such as CDWs.
Couplings driving the topological Mott transitions are also relevant to formations of a CDW, which
has not been satisfactorily discussed in the previous studies.
Since the emergence of the TMI in the honeycomb lattice
requires the Coulomb repulsion between the next nearest neighbor sites, the long-period CDW instability must be considered on equal footing,
which is not captured in the small-unit-cell mean-field ansatz employed in the previous studies.
Examining charge fluctuations with finite momentum over entire Brillouin zones is an alternative way to clarify the competitions
among TMIs and CDWs, as studied by employing functional renormalization group methods~\cite{Raghu,Scherer}.
However, first order thermal or quantum phase transitions not characterized by
diverging order-parameter fluctuations are hardly captured by such theoretical methods.
The most plausible symmetry breking competing with TMIs indeed occurs as a first order quantum phase transition as discussed later.

The quantum many-body fluctuations beyond the mean-field approximation severely affects the stability of the TMI.
The stability of the TMI and estimation of the critical value of interaction on the honeycomb lattice has mainly been considered by mean-field calculations
%\gr{[how about fRG ?]}
which can not treat the correlation effect satisfactorily.
%\sout{While in the limit of small electron correlation in the presence of quadratic band crossing, several studies have been done using renormalization group treatment\cite{Sun, Nandkishore} or first principle calculation\cite{Kitamura}
%, the band crossing point of the honeycomb lattice has finite Fermi velocity, and the relation between Fermi velocity and stability of the TMI should be revealed precisely.}
%\gr{[$\ast$ I guessed what Kurita-kun want to say and rewrote, but, the following sentences are not logical yet.]}
However, there exists a reliable limit where
the TMI becomes stable:
For infinitesimally small relevant Coulomb repulsions, the quadratic band crossing with
vanishing Fermi velocities
cause the leading instability toward the TMI,
as extensively examined by using perturbative renormalization group methods\cite{Sun, Nandkishore}.
However, examining the instabilities toward the TMI in Dirac semimetals
requires elaborate theoretical treatments.

%In this study, to estimate the stability of the TMI phase beyond the mean-field calculation, we use MVMC method\cite{Tahara, Kurita3} which will be presented in the following section.
In this study,
for clarification of the competitions among TMIs and other symmetry breakings, 
we first examine the long-period CDW at the level of mean-field approximation that
turns out to be much more stable compared to that of short period.
Indeed, this CDW severly competes the TMI on the honeycomb lattice.
The TMI on the honeycomb lattice studied in the literatures is consequently taken over by the CDW.

We, however, found a prescription to stabilize the TMIs on the honeycomb lattice:
By reducing the Fermi velocity of the Dirac cones, the TMI tends to be stabilized.
We examine the realization of the TMIs in the extended Hubbard model
on the honeycomb lattice by controlling the Fermi velocity and
employing
a variational Monte Carlo method\cite{Gros} with many variational
parameters\cite{Sorella},
multi-variable variational Monte Carlo (MVMC)\cite{Tahara, Kurita3},
together with the mean-field approximation.
Finally, we found that,
by suppressing the Fermi velocity to a tenth of that of the original honeycomb lattice,
the TMI emerges in an extended parameter region as a spontaneous symmetry breaking
even when we take many-body and quantum fluctuations into account.

This paper is organized as follows.
In section \ref{sec:Model and Method}, we introduce an extended Hubbard model and explain the order parameter of TMI.
We also introduce the MVMC method.
In section \ref{sec:Stability}, we first show how the long-range CDW becomes stable over the TMI phase in standard honeycomb lattice models.
Then we pursue the stabilization of TMI by modulating Fermi velocity at the Dirac cone at the mean-field level.
Finally we study by the MVMC method the effect of on-site Coulomb interaction  which was expected to unchange the stability of the TMI phase at the level of mean-field approximation.
Section \ref{sec:Dis} is devoted to proposal for realization of TMIs in real materials such as twisted bilayer graphene.  

\section{Model and Method}\label{sec:Model and Method}

\subsection{Extended Hubbard Model}
\begin{figure}[h]
\centering
\includegraphics[width=6.5cm]{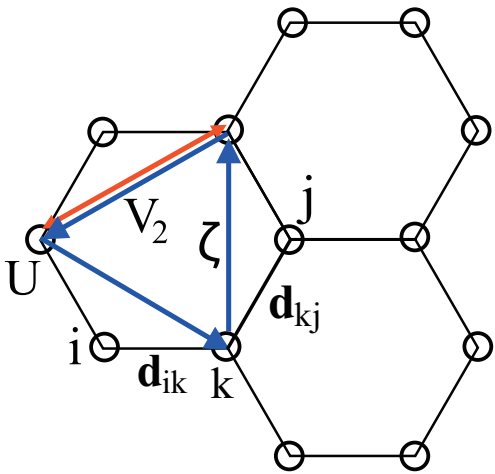}
\caption{
(Color online) Schematic picture of honeycomb lattice. 
Here, the on-site Coulomb interaction is denoted by $U$.
Next nearest neighbor interaction $V_2$ and the loop current $\zeta$ flowing between the next nearest neighbor are shown by red and blue arrows. 
}
\label{fig:honeycomb}
\end{figure}

In this section, we study ground states of an extended Hubbard model on the honeycomb lattice at half filling defined by
\begin{eqnarray}
	H &=& H_{0} + H_{\mathrm{SO}} + U\sum_{i}n_{i\uparrow}n_{i\downarrow} + \sum_{i,j}\frac{V_{ij}}{2}n_in_j,
\end{eqnarray}
where the single particle parts of $H$ are defined as
\begin{eqnarray}
	H_{0} &=& -\sum_{i,j}t_{ij} c_{i\sigma}^{\dagger}c_{j\sigma},
	\label{eq:H0}
\end{eqnarray}
and
\begin{eqnarray}
	H_{\mathrm{SO}} &=& i\lambda \sum_{\langle\langle i, j\rangle\rangle}\sum_{\alpha, \beta=\uparrow,\downarrow}
\left(\frac{\bm{d}_{ik}\times \bm{d}_{kj}}{|\bm{d}_{ik} \times \bm{d}_{kj}|}\cdot \bm{\sigma}
\right)_{\alpha\beta}c_{i\alpha}^{\dagger}c_{j\beta}\label{HSO}
\end{eqnarray}
is the spin-orbit interaction.
Here $c_{i\sigma}^{\dagger} \ (c_{i\sigma})$ is a creation (annihilation) operator for a $\sigma$- spin electron,
$n_{i} = n_{i\uparrow} + n_{i\downarrow}$ is an electron density operator, $t_{ij}$ represents the hopping of electrons between site $i$ and $j$, and $U(V_{ij})$ are on-site (off-site) Coulomb repulsion.
Bracket $\langle \langle i,j \rangle \rangle$ denotes the next-neighbor pair, $\lambda$ is the strength of the spin-orbit interaction
and $\bm{\sigma}=(\sigma_x,\sigma_y,\sigma_z)$ is the $S$=$1/2$-spin operator.
In Eq.(\ref{HSO}), the $k$-th site is in the middle of the next nearest neighboring pair $i$ and $j$ as shown in Fig. \ref{fig:honeycomb}, and $\bm{d}_{ij}$ is the vector from the site $i$ to $j$.

We start with the hopping matrix $t_{ij}$ in Eq.(\ref{eq:H0}) for the bond connecting a pair of the nearest-neighbor sites $\langle i,j\rangle$, 
\begin{eqnarray}
H_{0} =  H_{1} = -t_{1}\sum_{\langle i,j \rangle \sigma} c^{\dagger}_{i\sigma}c_{j\sigma}, \notag
\end{eqnarray}
as the simplest extended Hubbard model on the honeycomb lattice.
Later, we will examine the effect of third neighbor hoppings $t_3$.
We take $t_1$ as the unit of energy and set $t_1=1$ throughout the rest of this paper.
For the non-interacting limit, $U=V_{ij}=0$,
the system becomes a topological insulator when $\lambda$ is nonzero,
which is identical to the topological phase of the Kane-Mele model~\cite{Kane1,Kane2}.

For the off-site Coulomb repulsion, we mainly consider the second neighbor interaction ($V_{ij} = V_{2}$),
which is necessary for the emergence of the correlation-induced topological insulator~\cite{Raghu}.
The second neighbor Coulomb repulsions $V_2$ effectively generate the spin-orbit interactions, which are identical to $\lambda$,
and induce topological insulator phases even for $\lambda=0$.

We note that the Coulomb repulsion of on-site or the nearest neighbor site do not affect the stability of the TMI phase at the level of the mean-field approximation.
Indeed, our MVMC results show that this is essentially true beyond the level of mean-field approximation which will be discussed in the later section.
Therefore, for the moment, we focus only on the effect of $V_{2}$ for the consideration of interaction effects.

Here the TMI is the broken symmetry phase characterized by the order parameter
$\zeta$ defined by
\begin{eqnarray}
	\zeta&=& \frac{i}{2}\sum_{\alpha\beta}\langle c_{j\beta}^{\dagger}c_{i\alpha}\rangle_{\rm NNN}\left( \frac{\bm{d}_{ik}\times\bm{d}_{kj}}{|\bm{d}_{ik}\times\bm{d}_{kj}|} \cdot \bm{\sigma} \right)_{\beta\alpha}\label{eq:MFHoney2},
\end{eqnarray}
where the self-consistent mean fields for the second neighbor bonds are given by
\begin{eqnarray}
	\langle c_{j\beta}^{\dagger}c_{i\alpha}\rangle_{\rm NNN} &=& -i\zeta
\left(\frac{\bm{d}_{ik}\times\bm{d}_{kj}}{|\bm{d}_{ik}\times\bm{d}_{kj}|}\cdot\bm{\sigma}
\right)_{\alpha\beta}. \notag \\
%	\label{eq:MFHoney2}
\end{eqnarray}
Here, $\langle\cdots\rangle_{\rm NNN}$ denotes the expectation value for next nearest neighbor bonds and the order parameter $\zeta$ is physically interpreted as spin dependent loop currents flowing within a hexagons constituting the honeycomb lattice.

At the level of the mean-field approximation, this quantum phase transition is understood by decoupling two-body electron correlation term of the next nearest neighbor bond,
$\displaystyle V_2\sum_{\langle\langle i,j \rangle\rangle}n_i n_j$, as
\begin{eqnarray}
	&&n_{i\alpha}n_{i\beta} \rightarrow
	\langle n_{i\alpha} \rangle n_{j\beta} + n_{i\alpha} \langle n_{j\beta} \rangle
	- \langle n_{i\alpha} \rangle \langle n_{j\beta} \rangle
	\nonumber\\
	&-&
	\langle c^{\dagger}_{i\alpha}c_{j\beta} \rangle c^{\dagger}_{j\beta}c_{i\alpha} - c^{\dagger}_{i\alpha}c_{j\beta}
	\langle c^{\dagger}_{j\beta}c_{i\alpha} \rangle + \left| \langle c^{\dagger}_{i\alpha}c_{j\beta} \rangle \right|^{2}.
\end{eqnarray}
We also note that this phase transition to TMI, which is proposed not only on the honeycomb lattice but also in several other lattice models, belongs to an unconventional universality class, which depends on the dimension of the system and the dispersion of the electron band\cite{Kurita, Assaad}.
%\gr{[cite Kurita et al. and Assaad-Herbut]}.

\subsection{Multi-Variable Variational Monte-Carlo Method}

In this section, we pursue the topological Mott phase transition by employing 
the mean-field analysis and the variational Monte Carlo method.
For the latter method, we use a trial wave function of the
Gutzwiller-Jastrow form,
$|\Psi \rangle = \mathcal{P}_{G}\mathcal{P}_{J}|\Psi_{0}\rangle$ with a one body part,
\begin{eqnarray}
	|\Psi_{0}\rangle \equiv \left( \sum_{i,j=1}^{N_s}f_{ij}c_{i\uparrow}^{\dagger}c_{j\downarrow}^{\dagger}\right)^{N_{e}/2}|0\rangle,
\end{eqnarray}
where $f_{ij}$ is the variational parameters and $N_{e}$ is the number of the electrons in the system.
Though this form of the wave function restricts
itself to the Hilbert subspace with
the zero total $z$-component of $S$=$1/2$, $S_{z}=0$,
it can describe topological phases 
on the honeycomb lattice as long as we use complex variables for $f_{ij}$.
Here, $\mathcal{P}_{G}$ and $\mathcal{P}_{J}$ are the Gutzwiller and Jastrow factors defined as
\begin{eqnarray}
	\mathcal{P}_{G} = \exp\left(-\sum_{i}g_{i}n_{i\uparrow}n_{i\downarrow}\right),
\end{eqnarray}
and
\begin{eqnarray}
	\mathcal{P}_{J} = \exp\left(-\frac{1}{2}\sum_{i\neq j}v_{ij}n_{i}n_{j} \right),
\end{eqnarray}
respectively,
with which the effects of electron correlations are taken into account beyond the level of mean-field approximation.
%In this study, we set $f_{ij}$ as complex variables in order that the wave function can describe the topological insulator state.

The expectation value,
$\langle \Psi | H|\Psi \rangle/\langle\Psi | \Psi \rangle$, is minimized with respect to variational parameters,
$f_{ij}$, $g_i$, and $v_{ij}$,
by using the Monte-Carlo sampling and using the stochastic reconfiguration method by calculating gradient of the energy and the overlap matrix in the parameter space\cite{Sorella, Tahara, Kurita3}.
We optimize the parameters by typically 2000 stochastic reconfiguration steps.

\subsection{Thermodynamic Limit}

In the present implementation
of the variational Monte Carlo method with complex variables,
the feasible system size for the calculation is about up to 300, from which we speculate properties in the thermodynamic limit.
For this purpose, we perform the size extrapolation using the standard formula
\begin{eqnarray}
	\zeta_{L=\infty,\lambda=0} = \lim_{\lambda \rightarrow 0} \lim_{L\rightarrow \infty} \langle \zeta(L,\lambda) \rangle,\label{zeta_limit}
\end{eqnarray}
where $L$ is the linear dimension of the system size.
We note that the order of taking the limit in the right hand side of Eq.(\ref{zeta_limit}) is also important for the validity of the extrapolation,
which is known as the textbook prescription for the defining spontaneous symmetry breakings.
Another practical way to determine the spontaneous symmetry breakings is in principle the finite size scaling of the correlation for the order parameter.
However, the latter finite size scaling is not practically easy.
The correlation of $\zeta$ becomes 
too small because $\zeta$ itself is about the order of $0.01$ and the correlation becomes the order of $10^{-4}$ which becomes comparable to the statistical error of the Monte Carlo sampling. 

For the size extrapolation, we fit the data of the finite size calculations by a polynomial of $1/L^2$, that is, we assume the size dependence by
\begin{eqnarray}
	\zeta(L,\lambda) = \zeta(L=\infty,\lambda) + \frac{a(\lambda)}{L^2} + \frac{b(\lambda)}{L^4}.
\end{eqnarray}
The above assumption for the finite size scaling is based on a practical observation and an analogy to the finite size scaling in the spin wave theory~\cite{Huse}.
As a practical observation,
$\zeta(L,\lambda)$ for the limit $\lambda\rightarrow 0$ and $U=V=0$
is scaled by $1/L^2$. 
%\begin{eqnarray}
%	\zeta(L=\infty,\lambda) = \zeta_{L=\infty, \lambda=0} + c\lambda + d\lambda^2
%\end{eqnarray}

\section{Stability of Phases}
\label{sec:Stability}
We examine the ground states of the extended Hubbard model on the honeycomb lattice
by tuning the on-site Coulomb repulsion $U$ and the second neighbor Coulomb repulsion $V_2$.
Even for the parameter sets favorable to the TMIs that have been studied in the pioneering works on the TMI~\cite{Raghu},
we show that the TMI is not stabilized when we take into account other competing orders overlooked in the literature.
However, here, we reveal that, by tuning the Fermi velocities of the Dirac cones, the TMI is indeed stabilized.
\subsection{Charge Density Wave}
\begin{figure}[h]
\centering
\includegraphics[width=8.0cm]{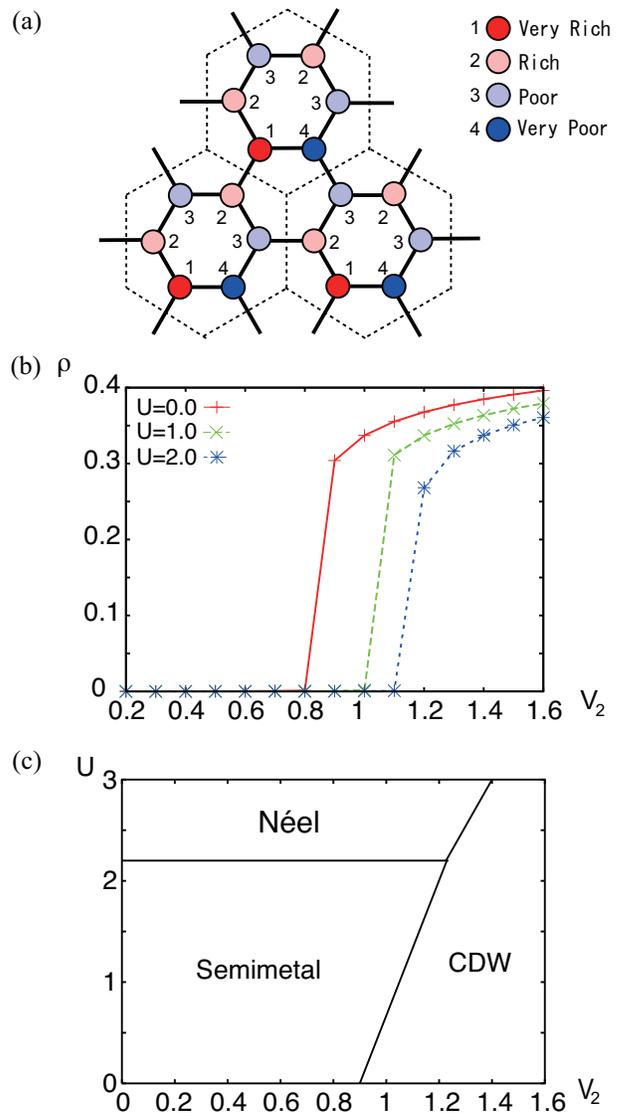}
\caption{
(Color online) (a) Schematic picture of CDW state on honeycomb lattice which is stabilized when $V_2$ is large.
Sites labelled by 1 (Red), 2 (light red), 3 (light blue) and 4 (blue) correspond to the sites where the electron densities are very rich, rich, poor and very poor, respectively.
(b) The growth of order parameter of the CDW state.
(c) Resulting phase diagram of honeycomb lattice for $U$ and $V_2$ at the level of mean-field approximation.
we do not find the region of stable TMI phase.
}
\label{fig:CDW6sub}
\end{figure}

In this section, we consider long-period CDW states and show that 6-sublattice order is stable when $V_2$ becomes dominat.
This state is schematically shown in Fig.\ref{fig:CDW6sub}(a), where the electron density
per site is disproportionated into four inequivalent values.
When we pick up very rich sites or very poor sites,
they constitute triangular lattices.
The moderately rich or poor sites
constitute the honeycomb lattices.

Figure \ref{fig:CDW6sub}(b) shows the growth of the order parameters of the 6-sublattice
CDW state for several different parameters of $U$.
In this calculation, we have used the mean-field approximation, where we defined the mean field by
\begin{eqnarray}
	\rho &=& \frac{1}{6}\sum_{n=1}^{6}|\rho_{n}|,
\end{eqnarray}
where $\rho_n$ is defined through
\begin{eqnarray}
	\langle c^{\dagger}_{\bm{r}n\sigma}c_{\bm{r}n\sigma} \rangle &=& \frac{1}{2} + \rho_{n} \quad (n = 1, \cdots, 6).
\end{eqnarray}
Here, the condition of half filling is satisfied by the following identity,
\begin{eqnarray}
	\sum_{n=1}^{6}\rho_{n} = 0.
\end{eqnarray}

As can be seen from
Fig.\ref{fig:CDW6sub}(b),
% the figure, 
the order parameter grows above $V_{2}=1.1$ even if we set $U=2.0$.
This is a quite serious problem for the stabilization of the TMI
since the critical value of $V_2$ for the TMI at the mean-field approximation is also about $V_{2}=1.2$. In addition, the energy gain due to the formation of
the topological Mott order is much smaller than that for the CDW state.
When we consider larger values of $U$, then the system becomes an antiferromagnetic state.
Furthermore, the mean-field approximation often overestimates the ordered phase, because it does not take into account fluctuation effects seriously.
Actually, by using the MVMC method, we could not find the region where the topological insulator becomes the ground state in the parameter space of $U$ and $V_{2}$.
The resulting phase diagram by the mean-field approximation is shown in Fig. \ref{fig:CDW6sub} (c).
There the antiferromagnetic phase is denoted as N\'eel,
%N\'eel represents the antiferromagnetic state 
where the spin on different sites of the bipartite aligns in the anti-parallel direction.

\subsection{Modulation of Fermi Velocity}

\begin{figure}
\centering
\includegraphics[width=8.5cm]{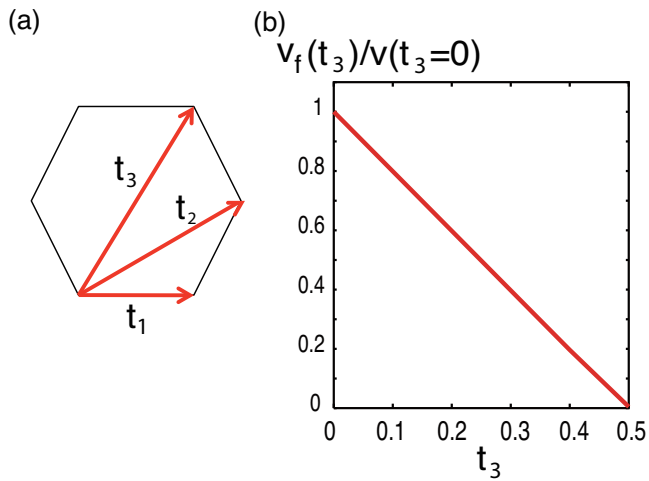}
\caption{
(Color online)
(a) Nearest, next nearest, and third neighbor 
hoppings.
(b) Relation between $t_{3}$ and Fermi velocity $v_{f}$ at Dirac point.
Here, $v_{f}$ is shown by the ratio to the Fermi velocity at $t_{3}=0$.
}
\label{fig:t3}
\end{figure}

As examined above, we found that the CDW state dominates and could not find parameter regions where the TMI is stable.
However, we find that the TMI becomes stable by modulating the Fermi velocity at the Dirac cones
% of the
% electron band. 
in the electronic band dispersion
Actually it is confirmed that, when the Fermi velocity is 0 and then the Dirac cones change to the quadratic band crossing points, the phase transition
from a zero-gap semiconductors to a TMI occurs with infinitesimal Coulomb repulsions~\cite{Sun}.
For the honeycomb lattice, it is possible to change
Dirac semimetals to quadratic band crossings by introducing the third neighbor hopping $t_3$ (schematically shown in Fig.\ref{fig:t3} (a)).
Then the part of the Hamiltonian $H_0$ is replaced with
\begin{eqnarray}
H_0 = H_1 + H_3,
%	H &=& H_{1} + H_{3} + U\sum_{i}n_{i\uparrow}n_{i\downarrow} + \sum_{\langle \langle ij\rangle \rangle }\frac{V_{2}}{2}n_in_j \notag \\
%	H_{1} &=& -t_{1}\sum_{\langle i,j \rangle \sigma} c^{\dagger}_{i\sigma}c_{j\sigma}, \notag \\
%	H_{3} &=& -t_{3}\sum_{\langle i,j \rangle_{TNN} \sigma} c^{\dagger}_{i\sigma}c_{j\sigma},
	\label{Eq.2}
\end{eqnarray}
where the third neighbor hoppings are given as
\begin{eqnarray}
H_3=-t_{3}\sum_{\langle i,j \rangle_{\rm TNN} \sigma} c^{\dagger}_{i\sigma}c_{j\sigma}.
\end{eqnarray}
Here the TNN stands for the third neighbor.
We find that the Fermi velocity linearly decreases by introducing $t_3$ and becomes $0$ at $t_{3}=0.5t_{1}$ as
$v_{f} = v_{f}(t_{3}=0)\times (1-2t_{3}/t_{1})$, which is also
shown in Fig.\ref{fig:t3} (b).
Though tuning nominal value of $t_{3}$ in the graphene is difficult, tuning of the Fermi velocity at the Dirac cone has been proposed in bilayer graphene by changing the relative orientation angle between two layers\cite{Morell}, which is effectively equivalent to the tuning of $t_{3}$.

\begin{figure}
\centering
\includegraphics[width=8.5cm]{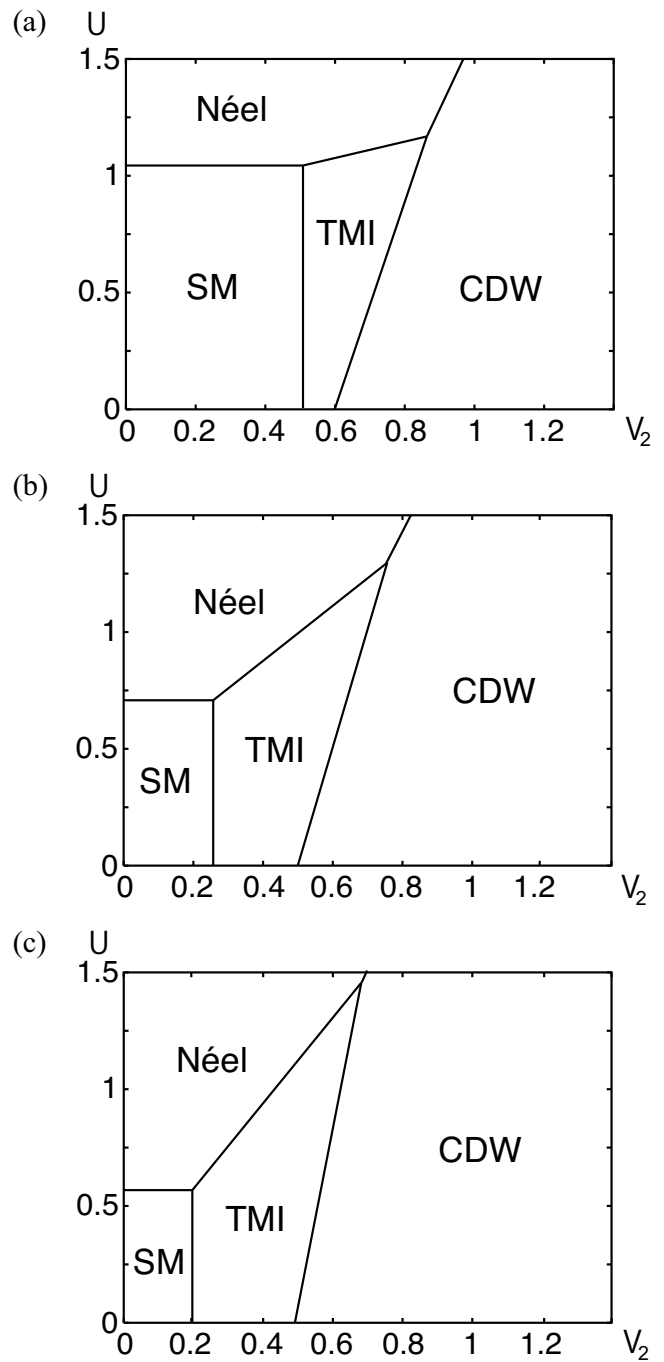}
\caption{
Phase diagram calculated by mean-field approximation for $t_{3}=$ (a) 0.3, (b) 0.4 and (c) 0.45.
SM denotes the semimetal.
}
\label{fig:MFphase}
\end{figure}

Figure \ref{fig:MFphase} shows the phase diagram calculated by the mean-field approximation for
$t_3=0.3$, $0.4$, and $0.45$.
% different values of $t_{3}$.
Compared to the N\'eel ordered phase and the TMI, the CDW phase is not largely affected by the Fermi velocity.
Therefore, the region of TMI recovers.

\subsection{Topological Mott insulator Studied by Multi-Variable Variational Monte-Carlo Method}\label{TMI_VMC}
Next, we examine the stability by using the MVMC method.
In our MVMC method, we impose the translational symmetry on the variational parameters to reduce the computational cost, as
\begin{eqnarray}
	f_{{\bm{r}}{\bm{r}'}} = f_{{\bm{r}+\bm{R}}\ {\bm{r}'}+\bm{R}}.
\end{eqnarray}
Here,
$\bm{R}$ is taken as any Bravais vector of the 6 sublattice unit cell illustrated in Fig.\ref{fig:CDW6sub}(a), in order
to examine possible spontaneous symmetry breakings including the CDW shown in Fig.\ref{fig:CDW6sub}(a)
and the TMI on an equal footing.
Therefore the number of  the sites for each calculation is taken as
$L\times L \times 6$. 
We have calculated for $L=4,5,6,7$, where about 2500 variational parameters are used for the calculation of the largest size.
To perform the extrapolation to the small external filed limit, $\lambda \rightarrow 0$, we have also calculated for several different strengths of the spin-orbit interaction, $\lambda = 0.001, 0.0005, 0.0002$, and $0.0001$.

\begin{figure}
\centering
\includegraphics[width=8.5cm]{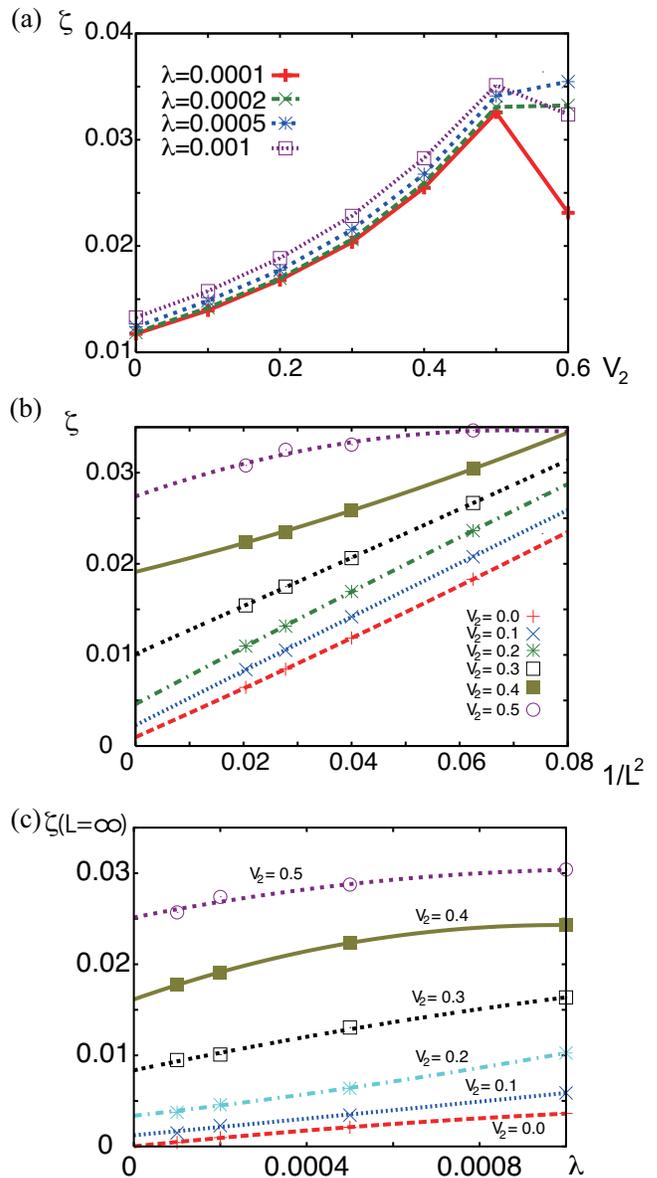}
\caption{
(Color online) Results calculated by the MVMC method. (a) MVMC results for several different values of $\lambda$ for $t_3 = 0.45$ at $L=5$.
(b) Size extrapolation at $\lambda = 0.0002$ for several different values of $V_{2}$. Here, four different sizes ($L=4,5,6,7$) are calculated.
Lines are results of quadratic fittings.
(c)Extrapolation of $\lambda$ is shown for different values of $V_{2}$. Lines are results of quadratic fitting.
}
\label{fig:VMCdata}
\end{figure}

Figure \ref{fig:VMCdata}(a) shows the numerical results for
the order parameter of the TMI at $t_{3}=0.45$ and $L=5$.
The sudden drops in $\zeta$ around $V_2=0.6$ signals the emergence of the 6-sublattice CDW state.
Indeed, when $V_{2}$ further increases, $\zeta$ vanishes.
This is physically quite natural because $\zeta$ is interpreted as the loop current and hard to be stabilized inside the CDW phase where electrons are locked at specific sites.

Same calculations are carried out for $L=4,6,7$, which is employed in the size
extrapolation of $\zeta$.
Figure \ref{fig:VMCdata}(b) shows the size extrapolation at $\lambda = 0.0002$ as a typical example.
Since the CDW becomes dominant at $V_2=0.6$,
the results for $V_2\lesssim 0.5$ are shown in Fig.\ref{fig:VMCdata}(b).
When $V_2$ is small, $\zeta$ empirically follows the size dependence $\zeta(\lambda)=\zeta(\lambda,L=\infty) + a/L^2$ with a constant $a$ as we see in Fig.5(b).
This extrapolation is performed for four different values of $\lambda$, and then we get the values for $\zeta(\lambda, L=\infty)$.
These data are shown in Fig. \ref{fig:VMCdata}(c), where the second extrapolation to $\lambda\rightarrow 0$ is performed.

\begin{figure}
\centering
\includegraphics[width=8.5cm]{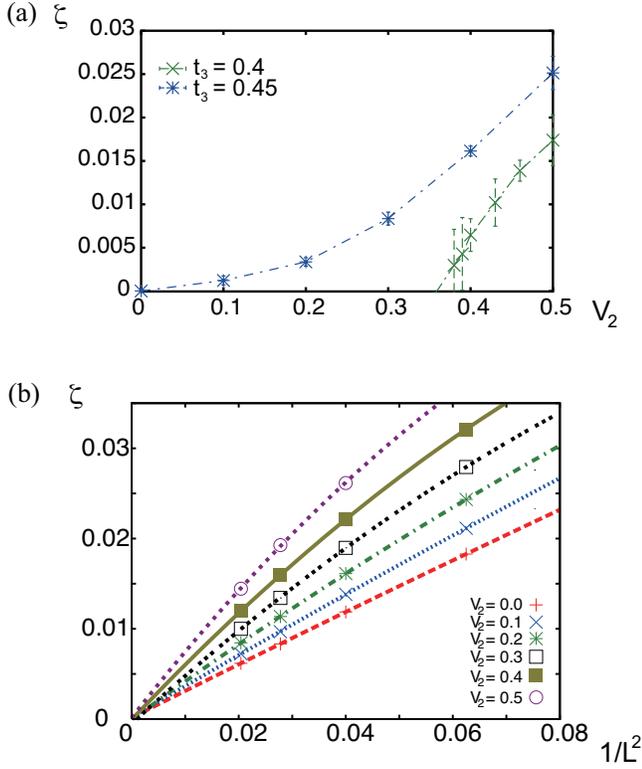}
\caption{(Color online) MVMC results. (a) Result of extrapolations for two different values of $t_{3}$ are shown.
(b)Size extrapolation at $t_{3}=0.3$ and $\lambda = 0.0002$ for several different values of $V_{2}$. Here, four different sizes ($L=4,5,6,7$) are calculated.
Lines are results of quadratic fittings.
}
\label{fig:VMCresult}
\end{figure}

\begin{figure}
\centering
\includegraphics[width=8.5cm]{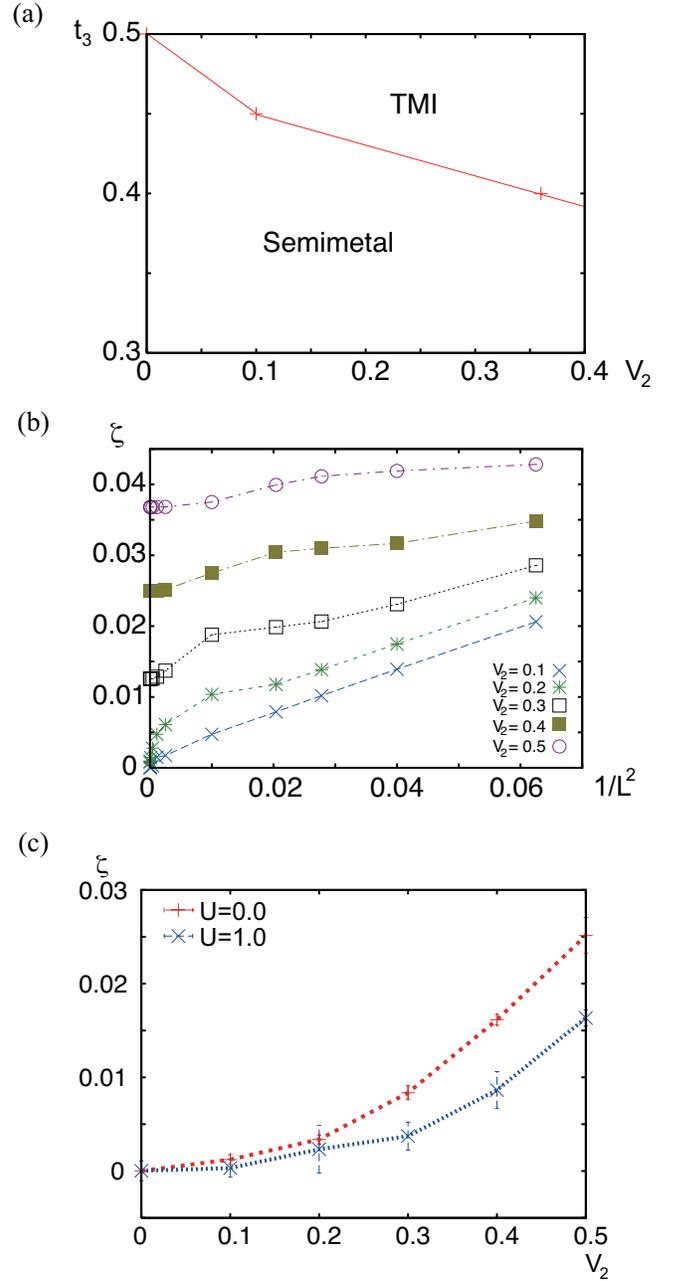}
\caption{(Color online) (a) Phase diagram of semimetal and the TMI with respect to $V_{2}$ and $t_{3}$ at $U=0$ obtained by MVMC calculations.
(b)Size dependence of mean-field calculation for several different values of $V_{2}$ at $t_{3}=0.45$.
 (c) $V_{2}$ dependence of order parameter of TMI for $U=0.0$ and $U=1.0$ at $t_{3} = 0.45$.
Decrease in $\zeta$ is observed by introducing $U$.
}
\label{fig:Ueffect}
\end{figure}

Final results for $\zeta (\lambda=0, L=\infty)$ are shown in Fig. \ref{fig:VMCresult}(a), where the results of $t_{3}=0.4$ are also shown.
The errorbars are defined from the errors of the extrapolation, where the largest error of the first extrapolation is added to the errors of the second extrapolation.
We note that the error bars arising from the statistical errors of the Monte Carlo sampling are much smaller.
Relatively large errorbars for small $V_{2}$ at $t_{3}=0.4$ is possiblly because of the existence of the critical point at a finite value of $V_{2}$, which is about $V_{2c}\sim 0.36$.
When $V_{2}$ exceeds this critical value and $\zeta$ in the thermodynamic limit remains nonzero, the error becomes smaller as can be seen from Fig. \ref{fig:VMCresult}(a).
%\gr{[$\ast$ I cannot understand the following sentence. It contradicts Fig.5(b):]}
For $t_{3}=0.45$, the order grows from small value of $V_{2} (\sim 0.1)$. 
However, at $t_{3}=0.4$, non-zero $\zeta$ can not be detected for small values of $V_{2}$.
Though the estimate of the universality class from these data is difficult, theoretically it is expected to belong to that of the Gross-Neveu model\cite{Herbut, Herbut2, Assaad}, and our result does not contradict this criticality. 
%\gr{[$\ast$ I cannot understand the following sentence:]}
For $t_{3} = 0.3$, we do not find the value of $V_{2}$ where $\zeta$ in the thermodynamic limit remains nonzero, and therefore phase transitions is not expected.
This is shown in Fig. \ref{fig:VMCresult}(b), where the size extrapolation at $\lambda = 0.0002$ is shown.
There, $\zeta$ becomes $0$ at $L \rightarrow \infty$ for all $V_{2}$, which is completely different from the behavior at $t_{3}=0.4$ and $t_{3}=0.45$.
The resulting phase diagram is shown in Fig. \ref{fig:Ueffect}(a).
Although we expect that the MVMC results show larger critical values $V_2=V_{2c}$ for the transition
at $t_3=0.45$, the estimated results indicate $V_{2c}$ slightly smaller than the mean-field results shown in Fig.\ref{fig:MFphase}. The reason that $V_{2c}$ becomes smaller %than that of the mean-field approximation 
at $t_3=0.45$ in the MVMC results 
is probably an artifact arising from a peculiar size dependence near the essential singularity at $t_{3}=0.5$,
as we see even in the mean-field calculation shown in Fig. \ref{fig:Ueffect}(b) where, the size dependence in the mean-field calculation are shown. The possible errors in in the estimate of $\zeta$ is as large as $0.005$ and the resultant errors in the estimate of $V_{2c}$ is around 0.1.  Therefore, the stability of the TMI phase over the CDW and N\'eel phases in the region $0.2 < V_2 <0.5$ for $t_3=0.45$ is robust. Here we note that the boundary between the CDW and TMI phases around $V_2=0.5$ does not change when we take into account the quantum fluctuations (by calculating with  the VMC method), because it is a strong first-order transition. 
%Quantitative difference of $V_{2c}$ at $U=0$ obtained by the mean-field approximation and MVMC calculations is about $0.1$ which is not as large as the range of $V_{2}$ where TMI becomes stable.
%Actually, we expect that the error of $\zeta$ due to the size extrapolation is .

Now we discuss the effect of the on-site Coulomb repulsion.
Though the onsite Coulomb repulsion does not affect the value of $\zeta$ and therefore stability of the TMI in the mean-field approximation, our MVMC result shows that increasing $U$ decreases the value of $\zeta$ if $V_{2}$ is fixed as shown in Fig. \ref{fig:Ueffect}(c).
While the increasing $U$ quantitatively decreases the value of $\zeta$, its effect does not destroy the stability of the TMI phase at $V_2>0.3$ at least if $U\sim t$. It also suppresses the emergence of the CDW.
Therefore it may help to enlarge the region of the TMI phase.
The same effect is expected when we consider the Coulomb repulsion for the nearest neighbor sites $V_{1}$.
That is, it decreases the value of $\zeta$ beyond the level of the mean-field approximation but does not essentially affect the phase transition.
However, we also note $V_{1}$ may cause another type of CDW, and stabilization of TMI should be examined against this CDW phase when $V_{1}$ is large.
%\gr{[$\ast$ I cannot understand the following sentence:]}
%Our calculation suggests that the emergence requires Coulomb interaction of the next nearest neighbor and that of on-site weaken the growth of the order parameter.

As a qualitative difference from the mean-field approximation, we found that effect of $U$ decreases the value of the order parameter of the TMI.
In the case of the border between the semimetal and the N\'eel ordered phases, it is expected that $U_{c}$, namely the critical values for $U$ becomes larger by treating fluctuation effects carefully.
Furthermore, we expect that $U_{c}$ becomes a function of $V_{2}$, which may enhance the fluctuation of the N\'eel order and suppress the phase transition.
In our calculation, we did not find N\'eel ordered states or mixed states of TMI and N\'eel ordered states.
On the other hand, the CDW phase is expected to be much more stable against fluctuations and the mean-field solution gives a reasonably good description because of a large scale of the energy gain for the CDW phase in comparison to other phases as adequately shown even by the mean-field approximation.
At the boundary of the TMI to N\'eel ordered phases, the universality class may change as suggested in the Kane-Mele-Hubbard model\cite{Hohenadler}.

\section{Possible Realization of Correlation Induced Topological Insulator in Twisted Bilayer Graphene}
\label{sec:Dis}
Here, we discuss the realization of the TMI in the real solids.
A primary candidate of TMIs is graphene.
As a well-known fact, graphene is nothing but a two-dimensional honeycomb network of carbon atoms
and hosts Dirac electrons.
However, it is also a well-known fact that, in free-standing graphene and graphene on substrates,
significant single-particle excitation gaps have not been observed yet\cite{Castro_Neto}.
Below, we examine possible routes toward the realization of the TMI in graphene-related systems.

First of all, as already studied above, the suppression of the Fermi velocity of the Dirac electrons
is crucial for the stabilization of the TMIs.
As extensively studied in the literature~\cite{Trambly},
twisted bilayer graphene (TBLG) offers Dirac electrons with tunable Fermi velocities.
By choosing stacking procedures, the quadratic band crossing, in other words, the zero Fermi velocity limit, is also achieved,
which has been already observed in experiments\cite{Luican, Ohta}.

Next, we need to clarify competitions with any other possible symmetry breakings in the TBLG with the Fermi velocity smaller than that of graphene.
The suppressed Fermi velocities may possibly cause instabilities towards not only the TMIs but also other competing orders as discussed in this paper.
For clarification of the competition we need an {\it ab initio} estimation of effective Coulomb repulsions
which directly correspond to the Coulomb repulsions in the extended Hubbard model\cite{Wehling}.
The {\it ab initio} study on the effective Coulomb repulsions employs a many-body perturbation scheme called constrained random phase approximation (cRPA)\cite{Aryasetiawan}.
The cRPA estimation gives the following values for the Coulomb repulsions: The on-site and off-site Coulomb repulsions are given as $U/t_1\simeq 3.3$,
$V_1/t_1\simeq 2.0$, and $V_2/t_1\simeq 1.5$ with $t_1 \simeq 2.8$ eV for free-standing graphene.
If we neglect longer-ranged Coulomb repulsions, we expect the N\'eel or CDW orders by employing these cRPA estimates
of $U, V_1$, and $V_2$ in graphene and TBLG.
Therefore, the free-standing graphene and TBLG do not offer a suitable platform for the TMIs.

However, by choosing dielectric substrates, the strength of the Coulomb repulsions, $U/t_1, V_1/t_1$, and $V_2/t_1$, is suppressed
due to enhancement of dielectric constant 
%$\varepsilon>1$
as  $U/\varepsilon t_1, V_1/ \varepsilon t_1$, and $V_2/\varepsilon t_1$,
where $\varepsilon$ is defined by dielectric constants of each materials as $\varepsilon =(\varepsilon^{\mathrm{graphene}} + \varepsilon^{\mathrm{substrate}})/\varepsilon^{\mathrm{graphene}}$.
(Here, we ignored the possible reduction of the effective dielectric constants at small distances.)
Then, we may approach the parameter region, where TMIs become stable, as shown in Fig.\ref{fig:Ueffect}(b).
If we neglect $V_1$, dielectric substrates with $\varepsilon \gtrsim 3$ are enough to stabilize the TMIs.
Even when the nearest-neighbor Coulomb repulsion $V_1$ is taken into account, TMI is expected to be stable as long as $V_1$ is not strong.

In the above discussion, we neglected further neighbor Coulomb repulsions, namely,
the third neighbor Coulomb repulsions $V_3$, the fourth neighbor ones $V_4$, and so on.
To justify the above discussion, we need to screen the further neighbor Coulomb repulsions by utilizing dielectric
responses of the substrates and/or adatoms.
Here we note that the screening from atomic orbitals on the same and neighboring sites effectively reduces the Coulomb repulsions
as extensively studied by using cRPA\cite{Wehling}.
The relative strength of the on-site and second neighbor Coulomb repulsions, $V_2/U$, may also be controllable
by utilizing the adatoms, which are expected to efficiently screen the on-site
% or nearest-neighbor 
Coulomb repulsions
if the adatoms is located just on top of the carbon atoms.
% or nearest-neighbor carbon-carbon bonds, respectively.
If we combine the control of the $V_2/U$ with the suppression of the further-neighbor Coulomb repulsions,
the above discussion may be relevant.
By utilizing the adatoms, the nearest-neighbor Coulomb repulsions $V_1$ are also expected to be well-screened
by adatoms on nearest-neighbor carbon-carbon bonds. The suppression of $V_1$ is helpfull for suppressing
the CDWs competing with the TMI.

Finally, we estimate the single-particle excitations gap $\Delta_{\rm c}$ induced by the TMI.
The excitation gap is crucial for actual applications of the TMI as a spintronics platform.
If we set $U/t_1\sim 1$ and expect the TBLG with $t_1\sim 3$ eV, we obtain the excitation gap up to 0.1 eV,
where we use the mean-field estimation of the gap $\Delta_{\rm c}=3\sqrt{3}\zeta V_2$\cite{Kurita} with the value of the MVMC result for $\zeta$ and assumed this formula is valid in the presence of electron correlation.
The estimated gap scale is substantially larger than the room temperatures.
Our theoretical results support that topological insulators with such a large excitation gap $\Delta_{\rm c}\sim 0.1$ eV are possibly obtained by using abundant carbon atoms.

\section{Conclusion}

In this paper, we have studied the realization of the TMI phase for the electronic systems on a honeycomb lattice by using the mean-field calculation and MVMC method.
We found that the CDW of the 6 sublattice unit cell is much more stable than the previously estimated CDW with smaller unit cells for the simplest case where the electronic transfer is limited to the nearest neighbor pair. For the stabilization of the TMI we need to suppress the Fermi velocity at the Dirac point than the standard Dirac dispersion for the case with only the nearest neighbor transfer.
In the case of the honeycomb lattice, this is realized by introducing the third neighbor hopping $t_3$ and we have given quantitative criteria for the emergence of the TMI.
Related real material is a bilayer graphene where the Fermi velocity is tuned by changing the rotation angle between two parallel layers\cite{Morell}.
Actually, the quadratic band crossing is realized when the rotation angle is $0$ (known as the AB stacking bilayer graphene\cite{Vafek, Murray}), which is mimicked by $t_3/t_1=0.5$.
Since smaller values of Fermi velocity stabilizes the TMI at smaller values of $V_{2}$, %\gr{[$\ast$ small compared with what ?]} 
%compared to that when the Fermi velocity is large,
its effective control may offer a breakthrough in the realization of two dimensional TMIs.
We need further analyses for experimental methods of controlling the stability of  the TMIs and {\it ab initio} quantitative estimates of the stability for bilayer graphenes, which are intriguing future subjects of our study.

\section*{Acknowledgement}

The authors thank financial support by Grant-in-Aid for Scientific Research 
(No. 22340090), from MEXT, Japan.
The authors thank T. Misawa and D. Tahara for fruitful discussions.
A part of this
research was supported by the Strategic Programs for Innovative
Research (SPIRE), MEXT (grant number hp130007 and hp140215), and the Computational Materials Science
Initiative (CMSI), Japan.


\begin{thebibliography}{99}

\bibitem{Kane1}
C. L. Kane, and E. J. Mele, {Phys. Rev. Lett.} {\bf 95}, 226801 (2005).

\bibitem{Kane2}
C. L. Kane, and E. J. Mele, {Phys. Rev. Lett.} {\bf 95}, 146802 (2005).

\bibitem{Fu}
L. Fu, C. L. Kane, and E. J. Mele, {Phys. Rev. Lett.} {\bf 98}, 106803 (2007).

\bibitem{Moore}
J. E. Moore, and L. Balents, {Phys. Rev. B} {\bf 75}, 121306 (2007).

\bibitem{Roy}
R. Roy, {Phys. Rev. B} {\bf 79}, 195322 (2009).

\bibitem{Hasan}
M. Z. Hasan, and C. L. Kane, {Rev. Mod. Phys.} {\bf 82}, 3045 (2010).

\bibitem{Bernevig}
B. Andrei Bernevig, Taylor L. Hughes, and S.-C. Zhang, {Science} {\bf 314} 1757 (2006).

\bibitem{Konig}
M. K\"{o}nig, S. Wiedmann, C. Br\"{u}ne, A. Roth, H. Buhmann, L. W. Molenkamp, X.-L. Qi, and S.-C. Zhang {Science} {\bf 318} 766 (2007).

\bibitem{Hsieh}
D. Hsieh, D. Qian, L. Wray, Y. Xia, Y. S. Hor, R. J. Cava, M. Z. Hasan {Nature} {\bf 452} 970 (2008).

\bibitem{Raghu}
S. Raghu, X.-L. Qi, C. Honerkamp, and S.-C. Zhang {Phys. Rev. Lett.} {\bf 100} 156401 (2008).

\bibitem{Wen}
J. Wen, A. R\"{u}egg, C.-C. J. Wang, and G. A. Fiete {Phys. Rev. B} {\bf 82} 075125 (2010).

\bibitem{Zhang}
Y. Zhang, Y. Ran, and A. Vishwanath {Phys. Rev. B} {\bf 79} 245331 (2009).

\bibitem{Kurita2}
M. Kurita, Y. Yamaji, and M. Imada {J. Phys. Soc. Jpn} {\bf 80} 044708 (2011).

\bibitem{Kurita}
M. Kurita, Y. Yamaji, and M. Imada {Phys. Rev. B} {\bf 88} 115143 (2013).

\bibitem{Herbut3}
I. F. Herbut and L. Janssen,
Phys. Rev. Lett. {\bf 113}, 106401 (2014).

\bibitem{Kitamura}
S. Kitamura, N. Tsuji, H. Aoki, arxiv:1411.3345 (2014)

\bibitem{Scherer} M. M. Scherer, S. Uebelacker, C. Honerkamp, Phys. Rev. B {\bf 85}, 235408 (2012). 

\bibitem{Sun}
K. Sun, H. Yao, E. Fradkin, and S. A. Kivelson, {Phys. Rev. Lett.} {\bf 103}, 046811 (2009).

\bibitem{Nandkishore}
R. Nandkishore, and L. Levitov, {Phys. Rev. B} {\bf 82}, 115124 (2010).

\bibitem{Gros}
C. Gros, {Ann. Phys} {\bf 189}, 53 (1989).

\bibitem{Sorella}
S. Sorella, Phys. Rev. Lett. {\bf 80}, 4558 (1998).

\bibitem{Tahara}
D. Tahara, and M. Imada, {J. Phys. Soc. Jpn} {\bf 77}, 114701 (2008).

\bibitem{Kurita3}
M. Kurita, Y. Yamaji, Satoshi Morita, and M. Imada, {Phys. Rev. B} {\bf 92}, 035122 (2015).

\bibitem{Huse}
D. A. Huse, Phys. Rev. B {\bf 37}, 2380 (1988).

\bibitem{Morell}
E. S. Morell, J. D. Correa, P. Vargas, M. Pacheco, and Z. Barticevic, Phys. Rev. B {\bf 82}, 121407(R) (2010).

\bibitem{Herbut}
I. F. Herbut, Phys. Rev. Lett. {\bf 97}, 146401 (2006).

\bibitem{Herbut2}
I. F. Herbut, V. Juri\v{c}i\'{c}, and O. Vafek, Phys. Rev. B {\bf 80}, 075432 (2009).

\bibitem{Assaad}
F. F. Assaad, and I. F. Herbut, Phys. Rev. X {\bf 3}, 031010 (2013).

\bibitem{Hohenadler}
M. Hohenadler, Z. Y. Meng, T. C. Lang, S. Wessel, A. Muramatsu, and F. F. Assaad, Phys. Rev. B {\bf 85}, 115132 (2012).

\bibitem{Vafek}
O. Vafek, and K. Yang, Phys. Rev. B {\bf 81} 041401(R) (2010).

\bibitem{Murray}
J. M. Murray, and O. Vafek, Phys. Rev. B{\bf 89} 201110(R) (2014).

\bibitem{Castro_Neto}
A. H. Castro Neto, F. Guinea, N. M. R. Peres, K. S. Novoselov, and A. K. Geim,
Rev. Mod. Phys. {\bf 81}, 109 (2009).

\bibitem{Trambly}
G. Trambly de Laissardi\`ere, D. Mayou, and L. Magaud,
Nano Lett. {\bf 10}, 804 (2010).

\bibitem{Ohta}
T. Ohta, A, Bostwick, T. Seyller, K. Hom, E. Rotenberg
Science {\bf 313}, 951 (2006)

\bibitem{Luican}
A. Luican, G. Li, A. Reina, J.Kong, R. R. Nair, K. S. Novoselov, A. K. Geim, and E. Y. Andrei
Phys. Rev. Lett. {\bf 106}, 126802 (2011)


\bibitem{Wehling}
T. O. Wehling,
E. \ifmmode \mbox{\c{S}}\else \c{S}\fi{}a\ifmmode \mbox{\c{s}}\else \c{s}\fi{}\ifmmode \imath \else \i \fi{}o\ifmmode \breve{g}\else \u{g}\fi{}lu,
C. Friedrich, A. I. Lichtenstein, M. I. Katsnelson, and S. Bl\"ugel,
Phys. Rev. Lett. {\bf 106}, 236805 (2011).

\bibitem{Aryasetiawan}
F. Aryasetiawan, M. Imada, A. Georges, G. Kotliar, S. Biermann, and 
 A. I. Lichtenstein,
Phys. Rev. B {\bf 70}, 195104 (2004).


\end{thebibliography}
\end{document}